\documentclass[aps,prd,twocolumn,nofootinbib,preprintnumbers,amssymb,amsfonts,amsmath,superscriptaddress,showpacs]{revtex4-1}

\usepackage{graphicx}
\usepackage{bm}
\usepackage{hyperref}
\usepackage{color}
\usepackage{subfigure}
\usepackage{url}
\usepackage{epstopdf}

\def\simleq{\; \raise0.3ex\hbox{$<$\kern-0.75em \raise-1.1ex\hbox{$\sim$}}\; }
\def\simgeq{\; \raise0.3ex\hbox{$>$\kern-0.75em \raise-1.1ex\hbox{$\sim$}}\; }

\newcommand{\GeV}{{\rm GeV}}

\newcommand{\TeV}{{\rm TeV}}

\newcommand{\erg}{{\rm erg}}

\newcommand{\kpc}{{\rm kpc}}
\newcommand{\pc}{{\rm pc}}

\newcommand{\cm}{{\rm cm}}
\newcommand{\km}{{\rm km}}

\newcommand{\s}{{\rm s}}

\begin{document}

\title{PAMELA and AMS-02 $e^+$ and $e^-$ spectra \\ are reproduced by three-dimensional cosmic-ray modeling}

\author{Daniele Gaggero}
\email{dgaggero@sissa.it}
\affiliation{SISSA, via Bonomea 265, I-34136, Trieste, Italy}
\affiliation{INFN, sezione di Trieste, via Valerio 2, I-34127, Trieste, Italy}

\author{Luca Maccione}
\email{luca.maccione@lmu.de}
\affiliation{Ludwig-Maximilians-Universit\"{a}t, Theresienstra{\ss}e 37, D-80333 M\"{u}nchen, Germany}
\affiliation{Max-Planck-Institut f\"{u}r Physik (Werner Heisenberg Institut), F\"{o}hringer Ring 6, D-80805 M\"{u}nchen, Germany}

\author{Dario Grasso}
\email{dario.grasso@pi.infn.it}
\affiliation{INFN, sezione di Pisa, Largo B. Pontecorvo 3, I-56127 Pisa, Italy}

\author{Giuseppe Di Bernardo}
\email{giuseppe.dibernardo@physics.gu.se}
\affiliation{Department of Physics, University of Gothenburg, SE 412 96 Gothenburg, Sweden} 

\author{Carmelo Evoli}
\email{carmelo.evoli@desy.de}
\affiliation{{II.} Institut f\"ur Theoretische Physik, Universit\"{a}t Hamburg, Luruper Chaussee 149, D-22761 Hamburg, Germany}



\begin{abstract}
The PAMELA collaboration recently released the $e^+$ absolute spectrum between 1 and $300~\GeV$ in addition to the positron fraction and the $e^-$ spectrum previously measured in the same period.
We use the newly developed three-dimensional upgrade of the {\sc DRAGON} package to model those data. This code allows us to consider a realistic spiral arm source distribution in the Galaxy, which impacts the 
high-energy shape of the propagated spectra.
At low energy we treat solar modulation with the {\sc HelioProp} code and compare its results with those obtained using the usual force-field approximation.   
We show that all PAMELA data sets can be consistently, and accurately, described in terms of a standard background on top of which a charge symmetric $e^+ + e^-$ extra component with harder injection spectrum is added; this extra contribution is peaked at $\sim 1 - 10 ~\TeV$ and may originate from a diffuse population of sources located in the Galactic arms.
For the first time, we compute the energy required to sustain such a relevant positron flux in the Galaxy, finding that it is naturally compatible with an astrophysical origin.  
We considered several reference propagation setups; we find that models with a low (or null) reacceleration -- tuned against light nuclei data -- nicely describe both PAMELA leptonic and hadronic data with no need to introduce a low-energy break in the proton and helium spectra, as it would be required for high reacceleration models.  
We also compare our models with the preliminary $e^-$ and $e^+$ absolute spectra recently measured by AMS-02. We find that those data, differently from what is inferred from the positron fraction alone, favor a high energy cutoff $\sim 10 ~\TeV$ of the extra component if this is uniquely generated in the Galactic arms. A lower cutoff may be allowed if a relevant contribution from powerful $e^- +\ e^+$ nearby accelerators ({\rm e.g.} one or few pulsars) is invoked.
\end{abstract}

\maketitle
\section{Introduction}

The PAMELA collaboration recently published the results of its first orbital measurement of the absolute positron spectrum performed in the $1.5 - 300~\GeV$ energy range \cite{Adriani:2013uda}.   
The positron fraction (PF) $e^+/(e^- + e^+)$ measured in the same energy range has been also reported in that paper as an update of previous PAMELA results \cite{Adriani:2008zr,Adriani:2010ib}.  
PAMELA previously released the $e^-$ spectrum, between 1 and 625 GeV \cite{Adriani:2011xv}.  
All those data sets were taken during the same period (from July 2006 to December 2009), hence in a very low solar activity period, which is well suited to study the interstellar spectra. Having contemporary measurements of the electron and positron spectra by the same experiment over a wide energy range is a very important achievement as it allows us to reduce not only experimental systematics but also the uncertainties related to cosmic ray (CR) propagation in the heliosphere (solar modulation) and in the Galaxy. The same collaboration also measured the proton and helium spectra \cite{Adriani:2011cu,Adriani:2013as}: that piece of information is very relevant in this context since it is required to compute the secondary components of $e^+$ and $e^-$ interstellar fluxes that are produced by the interaction of those CR species with the interstellar medium.

PAMELA provided the first firm evidence of a rising PF above 10 GeV \cite{Adriani:2008zr}, which was confirmed by Fermi-LAT \cite{FermiLAT:2011ab} later on. 
More recently, the AMS-02 collaboration measured the same quantity with higher accuracy up to 350 GeV. Its results agree with PAMELA up to 250 GeV and suggest a flattening of the PF above that energy \cite{Aguilar:2013qda}. The AMS-02 collaboration also released preliminary data on electron, positron, and proton spectra and the boron-over-carbon ratio. 
 
In the standard scenario, only $e^-$ receive a primary contribution from astrophysical sources so that the observed $e^+$ flux should uniquely be originated by spallation processes in the interactions of CR protons and helium nuclei with the interstellar gas.      
That scenario, however, fails to reproduce the PF rise observed by PAMELA, Fermi-LAT, and AMS-02 above 10 GeV, for any of the propagation models which are validated against proton and light nuclei CR data. Hence, the presence of a positron primary component with a hard spectrum was invoked to explain this discrepancy. It was shown (see, e.g., Refs.~\cite{DiBernardo:2010is} and references therein) that -- if this {\em extra component} is charge symmetric and it has a hard source spectrum $\displaystyle N_{e^\pm}(E) \propto E^{- \gamma_{e^\pm}} \exp{\left(-E/E_{\rm cut}\right)}$ with  $\gamma_{e^\pm} \simeq 1.5$ and $E_{\rm cut} \simeq~1~\TeV$ -- it allows us to consistently reproduce the PF measured by PAMELA  \cite{Adriani:2008zr} and the $e^+ + e^-$ spectrum measured by Fermi-LAT \cite{Abdo:2009zk,Ackermann:2010ij} and H.E.S.S. \cite{Aharonian:2009ah}.

An impressive amount of work has been devoted to explain this excess either in terms of astrophysical mechanisms -- mainly \ (A) pair production in the pulsar magnetospheres 
and \ (B)  secondary production near the supernova remnant (SNR) shocks -- or invoking particle dark matter (DM) annihilation or decay. 
See, e.g., Refs.~\cite{Serpico:2011wg,Cirelli:2012tf} for comprehensive reviews.  
In spite of those efforts, the nature of such components is still an open problem. 

Possible features in the lepton spectra or a dipole anisotropy due to local sources (e.g., pulsars or SNRs) have been proposed as possible signatures which may validate or reject some of those interpretations (see, e.g., Ref.~\cite{Malyshev:2009tw}). 
A realistic modeling of those features, however, requires one to carefully account for the complex three-dimensional spatial distribution of CR sources because above 100 GeV the energy loss length of electrons and positrons is comparable with the inhomogeneity length scale of the source distribution.  
This may require large corrections to the spatial distribution and energy spectra of Galactic electrons and positrons with respect to the predictions of two-dimensional semi-analytical or numerical models which assume azimuthal symmetry. 
We addressed this issue by developing a three-dimensional upgrading of the {\sc DRAGON} code~\cite{Evoli:2008dv}.
Using this code to account for the spiral-arm pattern of supernova remnants inferred from astronomical data, we showed that several problems that two-dimensional models face when trying to describe experimental data are significantly ameliorated \cite{Gaggero:2013rya}. 

The absolute positron spectrum recently published by PAMELA provides an important test of the results obtained in Ref.~\cite{Gaggero:2013rya}, allowing us to improve our understanding of the model parameters, and, together with the preliminary AMS-02 results, it permits new insights on the nature of the extra component. 

\section{Electron and positron spectra with a Galactic extra component}\label{sec:only_arms}

In this section we extend the analysis performed in Ref.~\cite{Gaggero:2013rya}, considering a wider set of propagation models and using recently released B/C and positron data. 
 
CR propagation is treated with the three-dimensional version of the {\sc DRAGON} code \footnote{A three-dimensional and fully anisotropic (this feature is not used in this work) version of
{\sc DRAGON} is available at \url{http://www.dragonproject.org/} for download.}, 
assuming CR nuclei as well as primary electrons and positron sources to be distributed along Galaxy spiral arms using the same pattern adopted in Ref.~\cite{Gaggero:2013rya}.
In this work we approximate the regular magnetic field to be azimuthal symmetric so that only the perpendicular component of the diffusion coefficient ($D_\perp = D_z = D$) is relevant. 
For this quantity we assume the following rigidity and spatial dependence: $D(\rho,z) = \beta^\eta D_0 (\rho/\rho_0)^\delta~\exp{\left(z/z_t \right)}$. 

Here we only considered a scale height $z_t = 4~\kpc$, which is the value preferred on the basis of $^9{\rm Be}/ ^{10}{\rm Be}$ data, having shown in Ref.~\cite{DiBernardo:2012zu} that the $e^-$ and $e^+$ spectra have a relatively small dependence on this parameter. 

\begin{figure}[t]
\centering
\includegraphics[width=0.48\textwidth]{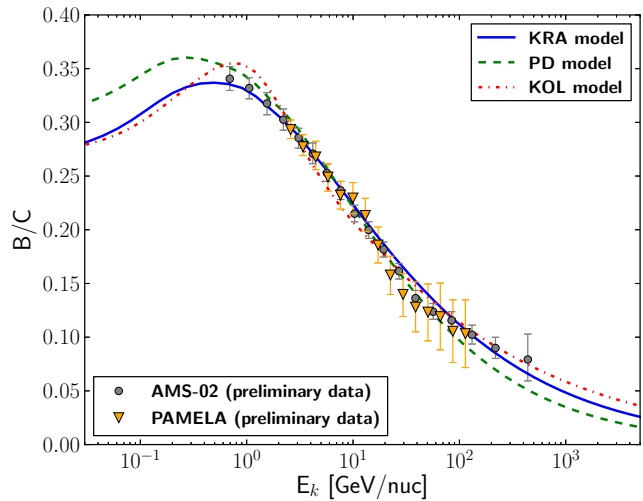}
\caption{The B/C ratio computed with DRAGON for the three propagation setups considered in this paper and modulated in the force-field approximation ($\Phi =  0.5~{\rm GV}$)  are compared with PAMELA \cite{BC_Pamela} and AMS-02 \cite{BC_AMS} preliminary experimental data. }
\label{fig:BC}
\end{figure}

\begin{figure}[t]
\centering
\includegraphics[width=0.48\textwidth]{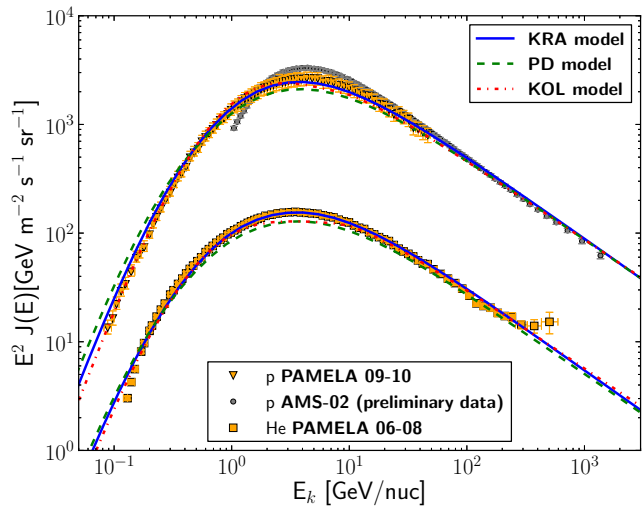}
\caption{The proton and helium spectra computed with DRAGON for the three propagation setups considered in this paper and modulated in the force-field approximation are compared with PAMELA and AMS-02 experimental data.  The local interstellar spectrum (LIS) is also shown. The PAMELA data reported in this and the following figures of this paper are extracted from the cosmic-ray database (CRDB) (http://lpsc.in2p3.fr/cosmic-rays-db/) \cite{Maurin:2013lwa}. }
\label{fig:protons}
\end{figure}

We consider three reference models defined by the parameters reported in Table~\ref{tab:models} which are tuned to reproduce PAMELA \cite{BC_Pamela} 
and AMS-02 \cite{BC_AMS} preliminary B/C data sets (see Fig. \ref{fig:BC}). 
What is even more relevant in this context is that these models also reproduce the proton and helium spectra measured by PAMELA both at high energies ($E > 80~\GeV$), where the effect of solar modulation is negligible, and at lower energies for a given choice of the solar modulation potential (see Fig.\ref{fig:protons}).  Here we disregard the spectral hardening at $E \sim 250~\GeV$ found by the PAMELA collaboration \cite{Adriani:2011cu} (and not confirmed by preliminary AMS-02 results) as we checked that its presence does not affect significantly our results.

\begin{table}[tbp]
\centering
  \begin{tabular}{|c|c|c|c|c|c|c|c|}
    \hline
      {\bf Model} & $\delta$  & $D_0$ & $v_A$  &  $\eta$ &  $\gamma(p)$ & $\gamma(e^-)$ & $\gamma(e^\pm)$  \\
   &  &  $(10^{28}\cm^{2}/\s)$ & $(\km/\s)$ & & & &  \\
    \hline
\hline    
	PD  &   0.5 & 2.7 & 0 & -0.40 & 2.50 & 1.7/2.5 & 1.7  \\
	KRA &  0.42 & 3.2 & 15 & -0.40 & 2.35 &1.8/2.5 & 1.7   \\
	KOL & 0.33 & 4.7 & 35 & 1 & 2.1/2.4 & 1.8/2.5  & 1.7   \\ 		
     	\hline	
  \end{tabular}
\caption{\label{tab:models} We report here the main parameters of the reference propagation setups used in this work. The KOL model has a break in the proton source spectrum $\gamma(p)$ at a rigidity of 11 GV.}
\label{tab:prop_para}
\end{table}

The three models are characterized by rather different values of the  Alfv\'en velocity $v_A$, which parametrize the reacceleration strength.
The plain diffusion (PD) propagation model ($v_A = 0$) used in this work is similar to that considered in Ref.~\cite{Gaggero:2013rya}. 
We also consider a weak reacceleration model with $\delta = 0.42$ (KRA) and a strong reacceleration model (KOL) with $\delta = 0.33$, which is similar to the best-fit model found in Ref.~\cite{Trotta:2010mx}. 
While the KOL model almost reproduces low-energy B/C data thanks to its strong reacceleration, it requires a break in the proton and helium source spectra to match their observed spectra. 
On the other hand, the PD and KRA adopt a single power-law injection spectrum for all nuclei at the cost of introducing an effective dependence of the diffusion coefficient on the particle velocity: 
$\eta < 0$.  Such behavior, however, can be motivated on theoretical grounds (see, e.g.,~Ref.~\cite{Evoli:2013lma}). 

For the electron background source spectrum, we adopt a broken power-law spectrum, with a break at $E = 4~\GeV$,  
as required to consistently match the synchrotron emission spectrum inferred from radio data and the $e^-$ spectrum measured by PAMELA (see Refs.~\cite{DiBernardo:2012zu}~and~\cite{Grasso:2013nyx}).  
As discussed in the introduction, the positron fraction data above 10 GeV require the presence of an $e^- + e^+$  extra component with a hard spectrum. We  assume the extra component to be charge symmetric and tune its source spectral index $\gamma(e^\pm)$  against the  PAMELA positron fraction data.
We consider two reference values of the extra component source spectrum cutoff energy: $E_{\rm cut} = 1$ and $10~\TeV$. 
The former is more suitable for pair production in pulsar wind nebulae (mechanism A mentioned in the introduction) while the latter is more natural for secondary production in SNRs (mechanism B). 

\begin{figure}[t]
\centering
\includegraphics[width=0.48\textwidth]{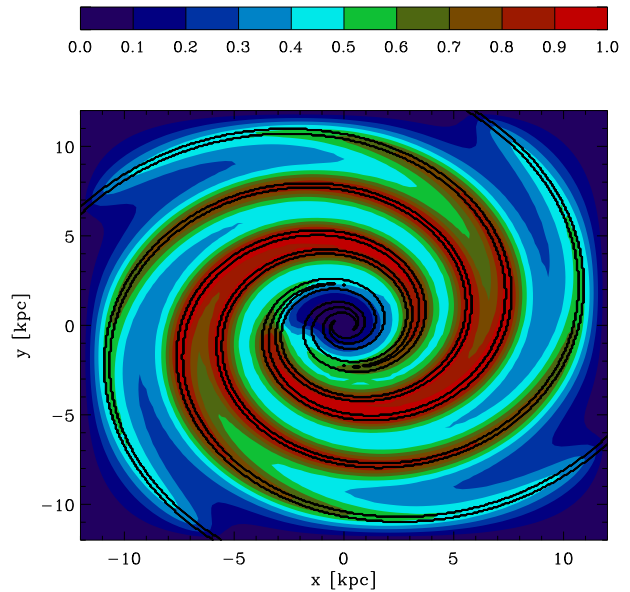}
\caption{The face-on view of the primary electron density at 100 GeV on the Galactic plane as computed with {\sc DRAGON} is represented (arbitrary units).}
\label{fig:contour_plot}
\end{figure}

\begin{figure}[h]
\centering
\includegraphics[width=0.48\textwidth]{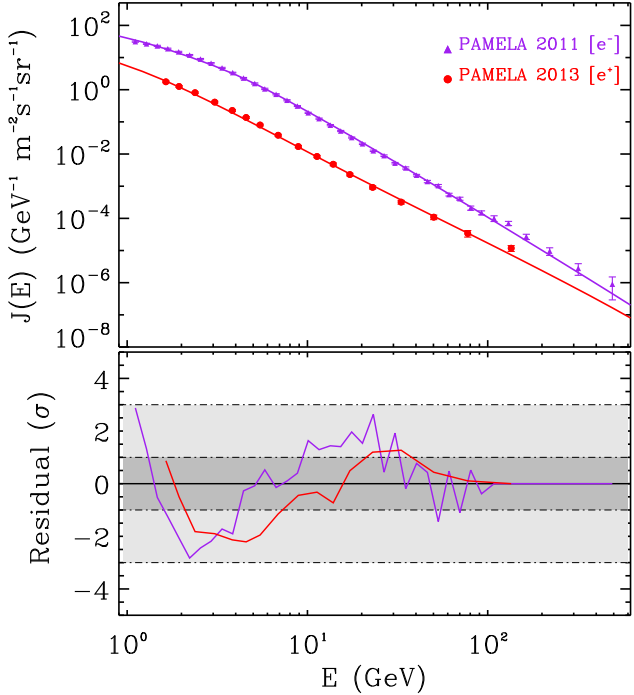}
\caption{The $e^-$ (purple solid) and $e^+$ (red solid) spectra propagated with the KRA setup, $E_{\rm cut} = 10~\TeV$ and modulated with {\sc HelioProp} are compared with PAMELA experimental data. The residuals of the model with respect to the data are shown in the lower panel. }
\label{fig:spectra}
\end{figure}

Similarly to what was done in Refs.~\cite{Gaggero:2013rya} and ~\cite{Grasso:2013nyx}, in this section we assume that the sources of this component have the same spiral-arm spatial distribution as that adopted for CR nuclei and for the electron background; this is consistent with both production mechanisms A and B mentioned above since both the pulsar and SNR populations are expected to be highly correlated with the spiral-arm structure.
The spatial distribution of the propagated high-energy electrons originated from a spiral source term is shown for illustrative purposes in Fig. \ref{fig:contour_plot}. 
%
%
\begin{figure}[t]
\centering
\includegraphics[width=0.43\textwidth]{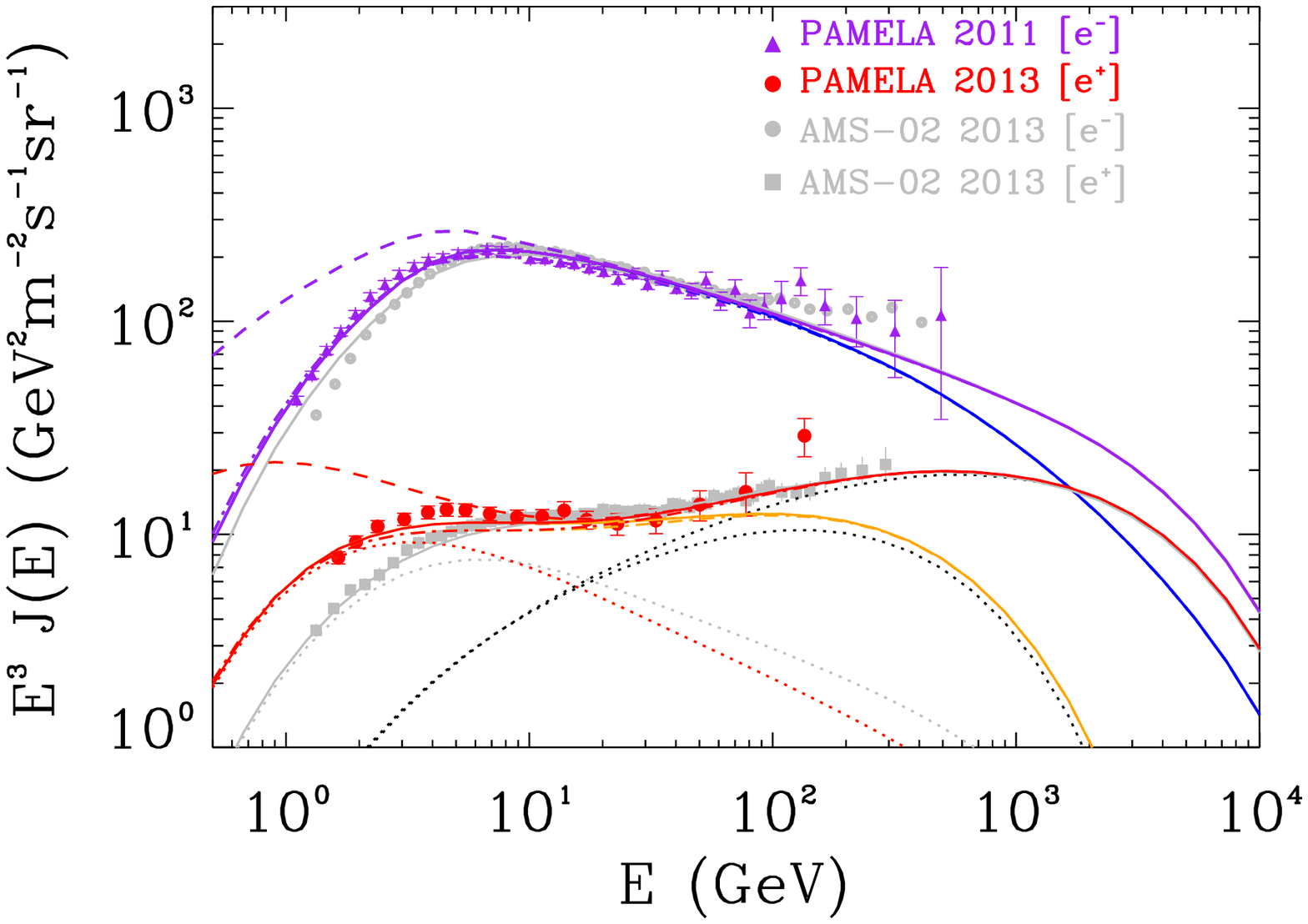}
\includegraphics[width=0.43\textwidth]{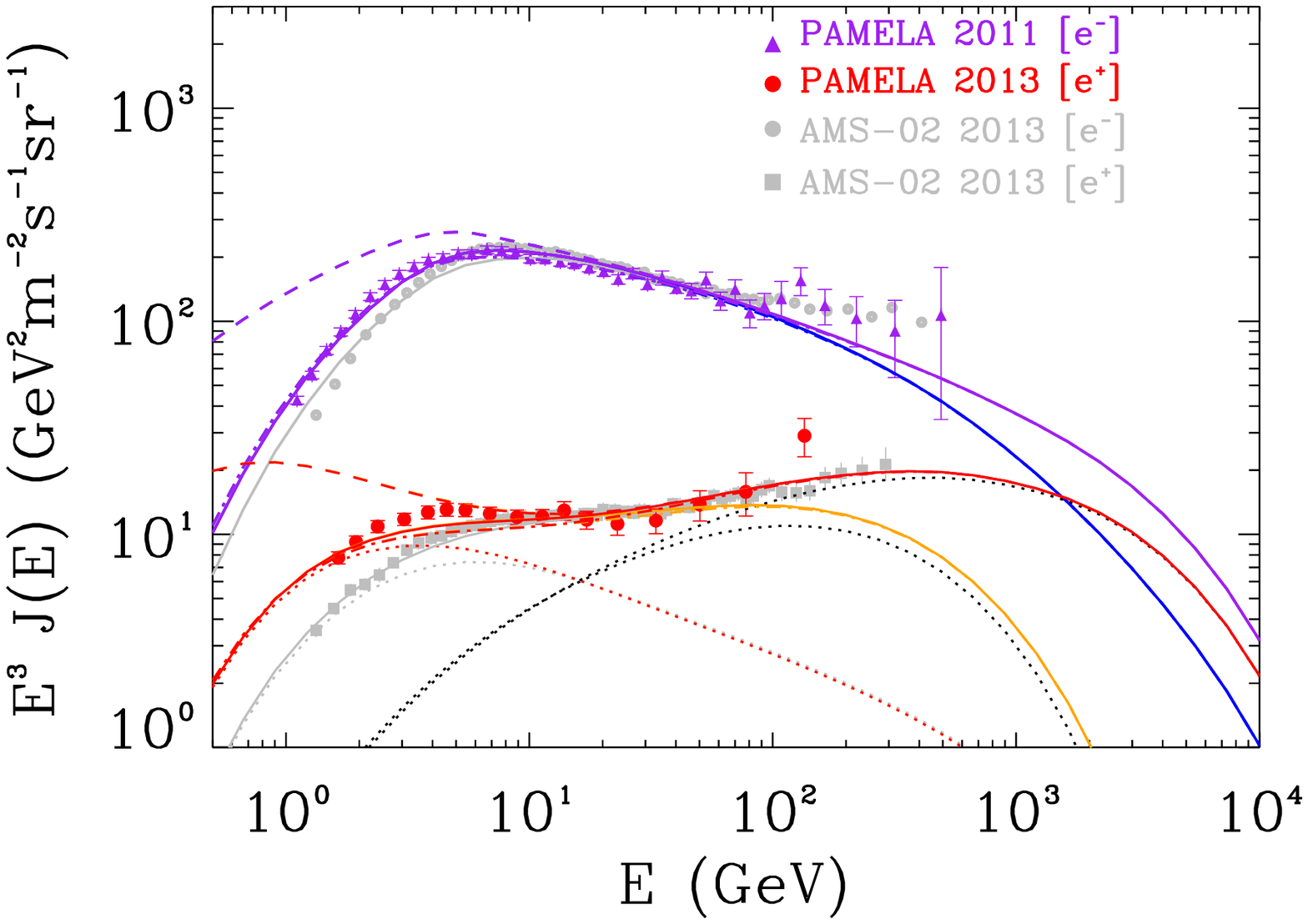}
\includegraphics[width=0.43\textwidth]{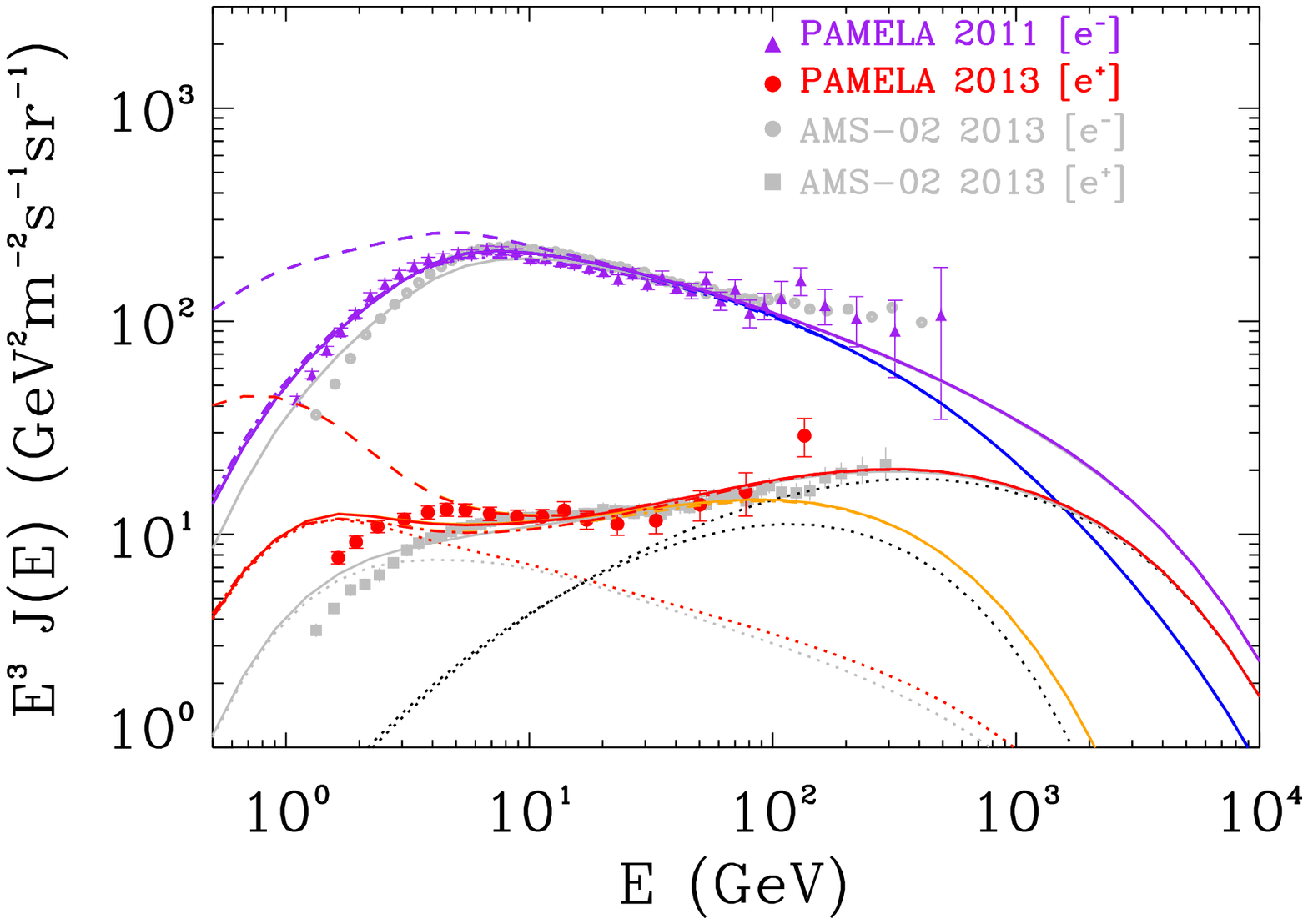}
\caption{
Red(orange), purple(blue) lines respectively represent the $e^+$, $e^-$ spectra computed for the extra component cutoff energy $E_{\rm cut} = \ 10 (1)\  \TeV$. 
Solid, dot-dashed and dashed lines represent charge dependent modulated, force-field modulated and interstellar (unmodulated) spectra respectively. The solid grey lines have been obtained with a modulation setup suited for AMS-02 data taking period.
Red and black dotted lines represent the modulated $e^+$ background and extra component respectively. From the top to bottom the three panels refer to PF, KRA and KOL models respectively. 
}
\label{fig:E3_spectra}
\end{figure}
\begin{figure}[t]
\centering
\includegraphics[width=0.46\textwidth]{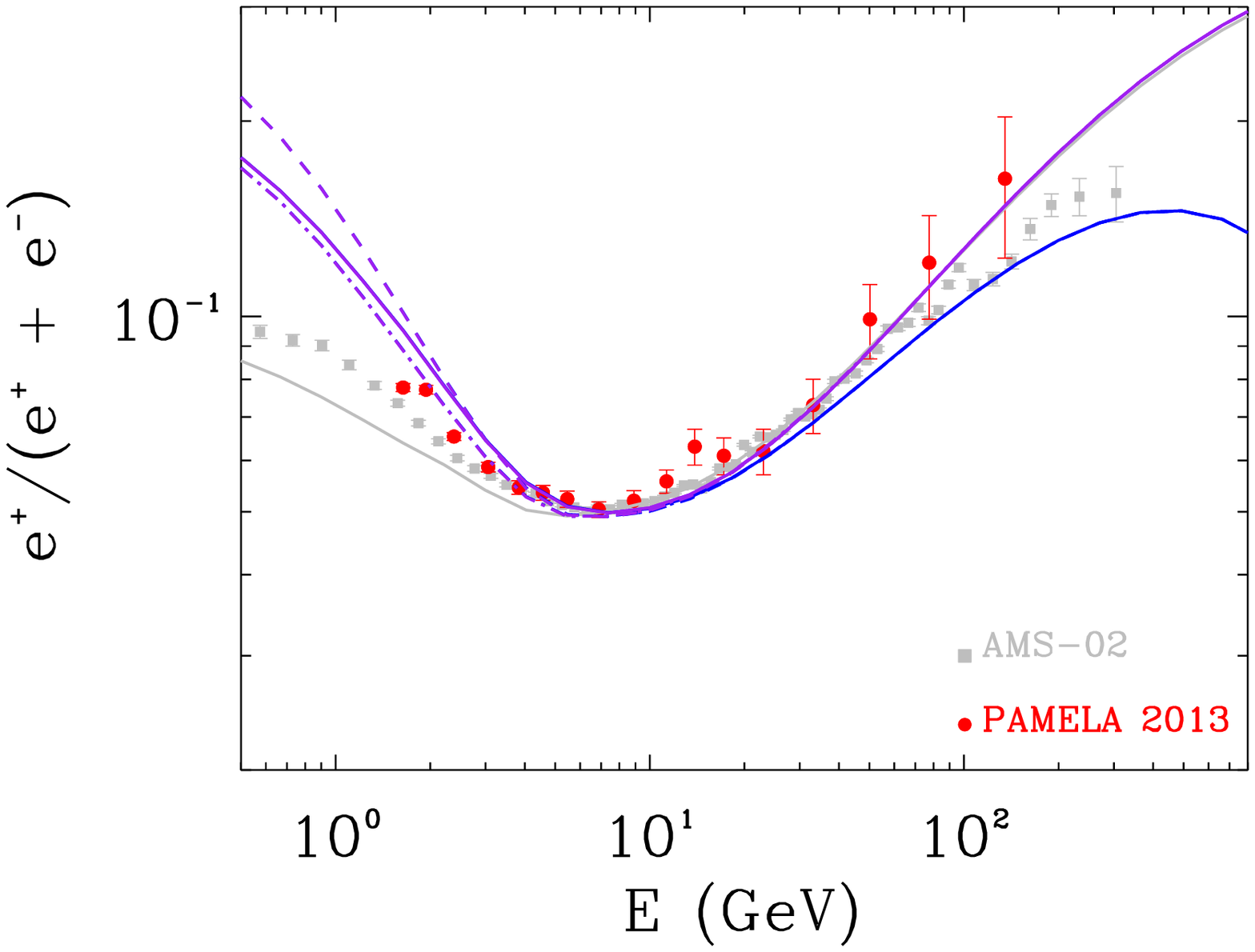}
\includegraphics[width=0.46\textwidth]{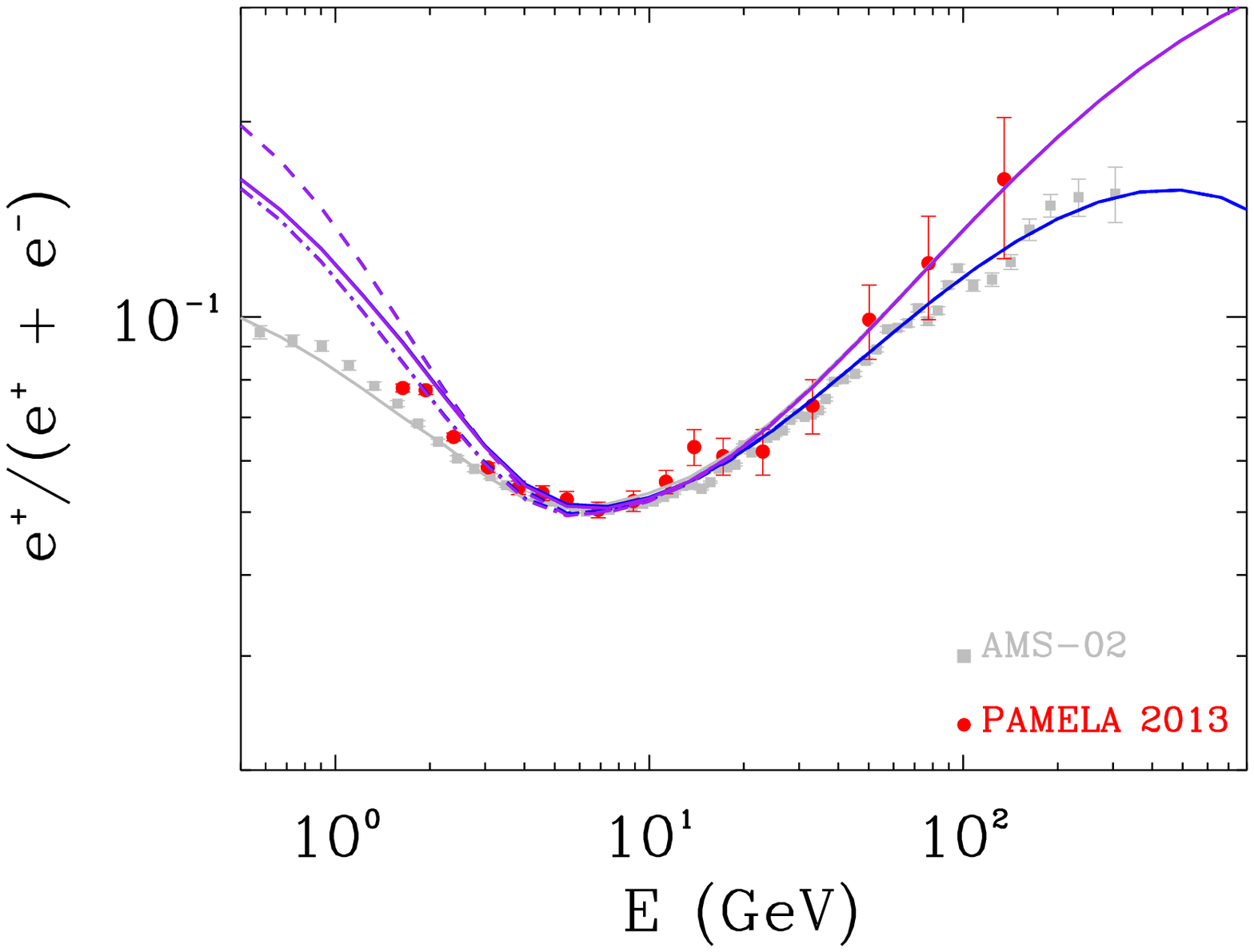}
\includegraphics[width=0.46\textwidth]{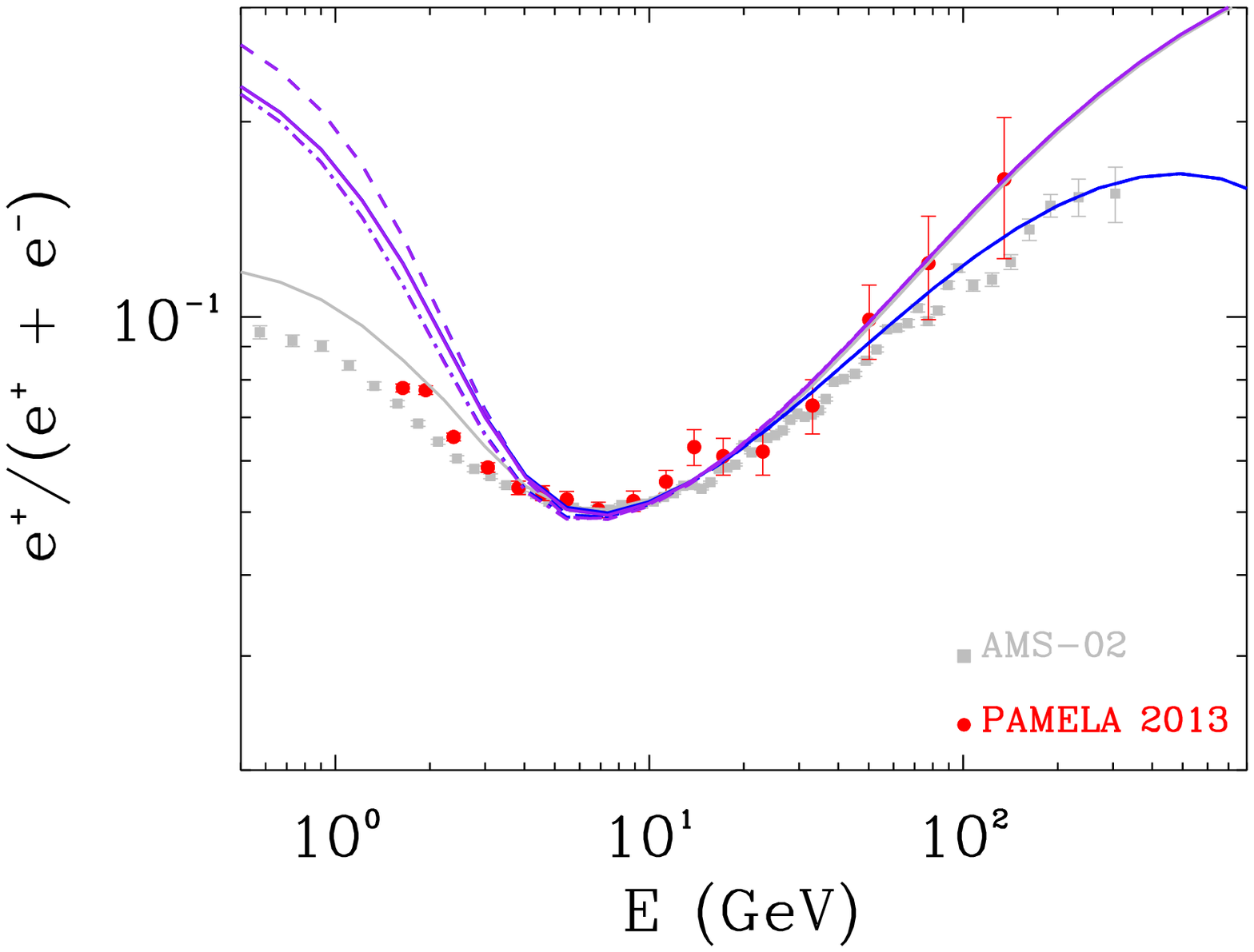}
\caption{The positron fraction corresponding to the models described in Sec.\ref{sec:only_arms} is compared with PAMELA and AMS-02 data. From the top to bottom the three panels have been obtained for the PF, KRA and KOL models respectively. The line notation is the same as that of Fig.\ref{fig:E3_spectra}.}
\label{fig:pos_frac}
\end{figure}

As first shown in Ref.~\cite{Gaggero:2013rya}, we notice the importance of this structured source term to reproduce the data using a more realistic primary injection spectrum: the enhanced energy losses due to the Sun being located in an interarm region, hence far from most sources, provide a further steepening which helps accommodate the extra component. 
We remark, however, that in order to match consistently all PAMELA data sets we need an injection spectrum [$\gamma(e^-) = -2.5$], which is still quite steeper than that expected from Fermi acceleration theory and inferred form radio observations of SNRs. 
A detailed investigation of the escape mechanism of the electrons from the sources should then be invoked to explain this discrepancy. 

Concerning solar modulation, we use both the widely used force-field, charge-independent, treatment \cite{Gleeson_1968ApJ} and the {\sc HelioProp} code \cite{Maccione:2012cu}, which solves the CR propagation equation in the Solar System accounting for charge-dependent drifts in the presence of a time-dependent current sheet (see also Ref.~\cite{Bobik:2011ig}).  
The second approach has been invoked to explain the discrepant PF low-energy data found during periods with opposite polarities of the heliospheric magnetic field and to match AMS-02 low-energy data.   
In {\sc HelioProp} a linear rigidity dependence of the scattering length parallel to the heliospheric magnetic field is assumed (Bohm diffusion): $\lambda_\parallel = \lambda_0 (\rho/1~\GeV) (B/B_\oplus)^{-1}$.  This scaling is different with respect to the one used in the interstellar medium and is motivated by the very high ratio between the turbulent and regular components of the magnetic field in the heliospheric environment.

We find that $\lambda_0 = 0.4$ astronomical units (AU) and the tilt angle current sheet  $\alpha = 10^\circ$,  which are well suited for the PAMELA data taking period,  provide a consistent description of the low-energy tail of $e^-$ and $e^+$ spectra measured by PAMELA for all models. 
From Fig.\ref{fig:E3_spectra} we see how both treatments (force field and charge dependent) give very similar results for the low solar activity period PAMELA was operating.
The main difference arises in the $e^+$ channel below 10 GeV where the charge dependent (CD) treatment allows a slightly better fit of PAMELA data. 
Both approaches consistently reproduce the $e^-$ and $e^+$ observed spectra. 
In the the force-field treatment, the values of the modulation potential required to reproduce the PAMELA lepton data for the PD, KRA and KOL models are, respectively, $\Phi = 0.35, 0.50, 0.25~{\rm GV}$. Noticeably, for our preferred model (KRA), the value of $\Phi$ coincides with that required to match the low-energy B/C data, and it is very close to that required for protons (see Figs.~\ref{fig:BC} and~\ref{fig:protons}). 
To properly compare our models also with AMS-02 data, we modulate our models with {\sc HelioProp} only
using a set of parameters more suitable for the active solar period during which they were taken:   $\lambda_0 = 0.22$ AU and $\alpha = 60^\circ$  (the force-feld approximation is less suited to describe solar modulation in such conditions).
 
From Figs.~\ref{fig:E3_spectra} and~\ref{fig:pos_frac} the reader can see as the lepton PAMELA data are consistently matched for the PD and KRA setups. In particular, the latter setup allows a remarkably good fit of PAMELA (and AMS-02 as well) low-energy data.  
The residuals of that model with respect to PAMELA data, which we show at the bottom of Fig.~\ref{fig:spectra}, are smaller than $2 \sigma$ over most of the considered energy range. 
The reduced $\chi^2$'s with respect to the $e^-$ ( $e^+$) data are $1.7(1.4);\ 1.9(1.8);\  4.1(9.3)$  for the PD, KRA, and KOL models, respectively. 
The KOL model is clearly disfavored by PAMELA $e^+$ data. The same would hold for any strong reacceleration model due to the bump it would produce in the $e^+$ spectrum below few GeV.
We find no significant difference between the $\chi^2$ computed with $E_{\rm cut} = \ 1$ and $10\  \TeV$ due to the large PAMELA errors. 
We notice, however, that the former case is disfavored by AMS-02 $e^-$ and $e^+$ preliminary data \cite{AMS_Schael}. 
Interestingly, this is opposite from what was inferred from the AMS-02 PF data alone (see the related discussion in Sec.\ref{sec:discussion}).

With DRAGON we also computed the injection power in the Galaxy required for protons, electron background, and extra component sources.
The results are as follows:3
\vskip0.2cm
\indent $W_{\rm tot}(p) \approx 10^{57}~\GeV/{\rm Myr}$; \\
\indent $W_{\rm tot}(e^-) \approx 10^{55}~\GeV/{\rm Myr}$; \\ 
\indent $W_{\rm extra}(e^\pm) \approx 10^{54}~\GeV/{\rm Myr}$.
\vskip0.2cm

We verified that these quantities are quite stable. In fact, the following are true:
\begin{enumerate}
\item Varying the propagation model, among those considered in this paper, results in a $30$ - $50$\% scatter of the powers.
\item Resizing the diffusive halo scale height in the interval $z_t = 1 - 16 \,{\rm kpc}$ results in fluctuations around $50$\%.
\item Variation of the extra component high-energy cutoff produces negligible changes.
\end{enumerate}

Therefore, the orders of magnitude estimates given above can be taken as good reference values for a wide class of scenarios.

Since the kinetic power released by supernovae in the interstellar medium (ISM) is $W_{\rm SNR} \approx 10^{58}~\GeV/{\rm Myr}$ (assuming a standard rate of about two explosions per century), these results are compatible with the common wisdom that CR protons and electrons are accelerated by SNRs. 
Interestingly, only a tiny efficiency $(10^{-4})$ would be required to power the extra component if this is originated by mechanism B (secondary production at SNR shock).
The Galactic pulsar population, however, can also provide the required energy budget. 
In this case, since the mean spin-down luminosity released by a mature pulsar in the form of dipole radiation is roughly 2 orders of magnitude smaller than the kinetic energy of a SNR (see, e.g., Eq.~(7) in Ref.~\cite{Serpico:2011wg}), the required pair creation efficiency would be $(\approx 1 - 10\%)$.
Interestingly this is significantly smaller than the values required in the scenario in which only few nearby pulsars are invoked to explain the positron excess \cite{Hooper:2008kg}.   

\section{Case of nearby sources}\label{sec:local_sources}

In the previous section, we assumed that CR electron and positron sources lay in the main spiral arms of the Galaxy. Since the closest arms are about 1 kpc away from the solar system, no nearby sources are present in that scenario. We know, however, that the solar system is not isolated but is part of a structure called the {\it local spur} or {\it local arm} (which includes also the Gould belt region); so several SNRs and pulsars are present in the solar neighborhood.
Few of these sources may significantly contribute to the observed $e^-$ and $e^+$ fluxes above 10 GeV (see Ref.~\cite{Serpico:2011wg} and references therein). 
In Sec.~\ref{sec:only_arms} we managed to show that such local components are not strictly necessary to get good fits of PAMELA data. 
However, a situation in which local sources reveal their presence in the spectral features on top of a Galactic diffuse component is worth being considered. 
Interestingly, when compared with our previous results (see Figs.~\ref{fig:E3_spectra} and~\ref{fig:pos_frac}) preliminary AMS-02 $e^+$ and $e^-$ data may already show some hints of such local contributions. In particular, the following are true:

\begin{itemize}
\item In the case of a high-energy extra component cutoff ($E_{\rm cut} = 10~\TeV$), we see that once the Galactic extra component is tuned to reproduce $e^+$ PAMELA and AMS-02 data, our Galactic model underestimates the $e^-$ spectrum measured by AMS-02 for $E > 100\ \GeV$, and, consequently, it overshoots the PF measured by that experiment. This may point to the presence of a local $e^-$ accelerator;
\item For a lower energy cutoff ($E_{\rm cut} = 1~\TeV$), the model reproducing PAMELA data underestimates both the $e^+$ and the $e^-$ spectra preliminary measured by AMS-02 (such discrepancies counterbalance in such a way as to match the AMS-02 PF). In this case the presence of a local $e^- + e^+$ accelerator may be required.   
\end{itemize}
 
Motivated by these arguments, in this section we consider two representative scenarios:

\begin{itemize}
\item A prominent nearby SNR, which for definiteness we assume to be Vela, injects $e^-$ only,
\item A few selected nearby pulsars, which, as motivated in previous studies~\cite{Hooper:2008kg,Grasso:2009ma}, we assume to be Monogem and Geminga, inject a charge symmetric $e^- +~ e^+$, component which provides an important contribution to the total flux.
\end{itemize}
 
While more local sources may contribute, previous analyses showed that their contribution should be subdominant with respect to those considered here \cite{DiBernardo:2010is,DiMauro:2014iia}.
In both cases we sum those local components to the $e^-$ and $e^+$ diffuse extra component originated in the Galactic arms and to the standard background. 

We treat particle propagation from the nearby sources to the solar system analytically, as described in Refs.~\cite{Grasso:2009ma,DiBernardo:2010is}, using the same propagation parameters and loss rate as those adopted in the numerical computation of the Galactic components. For definiteness in this section we consider only the KRA propagation setup. 

\subsection*{A. Adding a $e^-$ component from a nearby SNR}
 
Vela SNR with an age of  $1.1 \times 10^4~{\rm years}$ and a distance $d \simeq 294~\pc$ \cite{Green:2009qf} has been often proposed as a prominent local source of CR electrons in the TeV energy range. 
We assume the $e^-$ injection in the ISM to be instantaneous (which is justified by the fact that the SNR active time is considerably smaller than the propagation time to the Solar System).
For the injection spectrum energy dependence, we take $J_{\rm SNR}(e^-) \propto E^{-\gamma_{\rm SNR}} \exp{\left(- E/E_{\rm SNR}\right)}$.  
Here we assume that the $e^-$ injected by Vela have total energy  $E = 2.3 \times 10^{48}~\erg$, spectral index $\gamma_{\rm SNR} = 2.4$, and $E_{\rm SNR} = 5~\TeV$, which is in agreement with observations and Fermi acceleration theory.    

\begin{figure*}[t]
\centering
\includegraphics[width=0.49\textwidth]{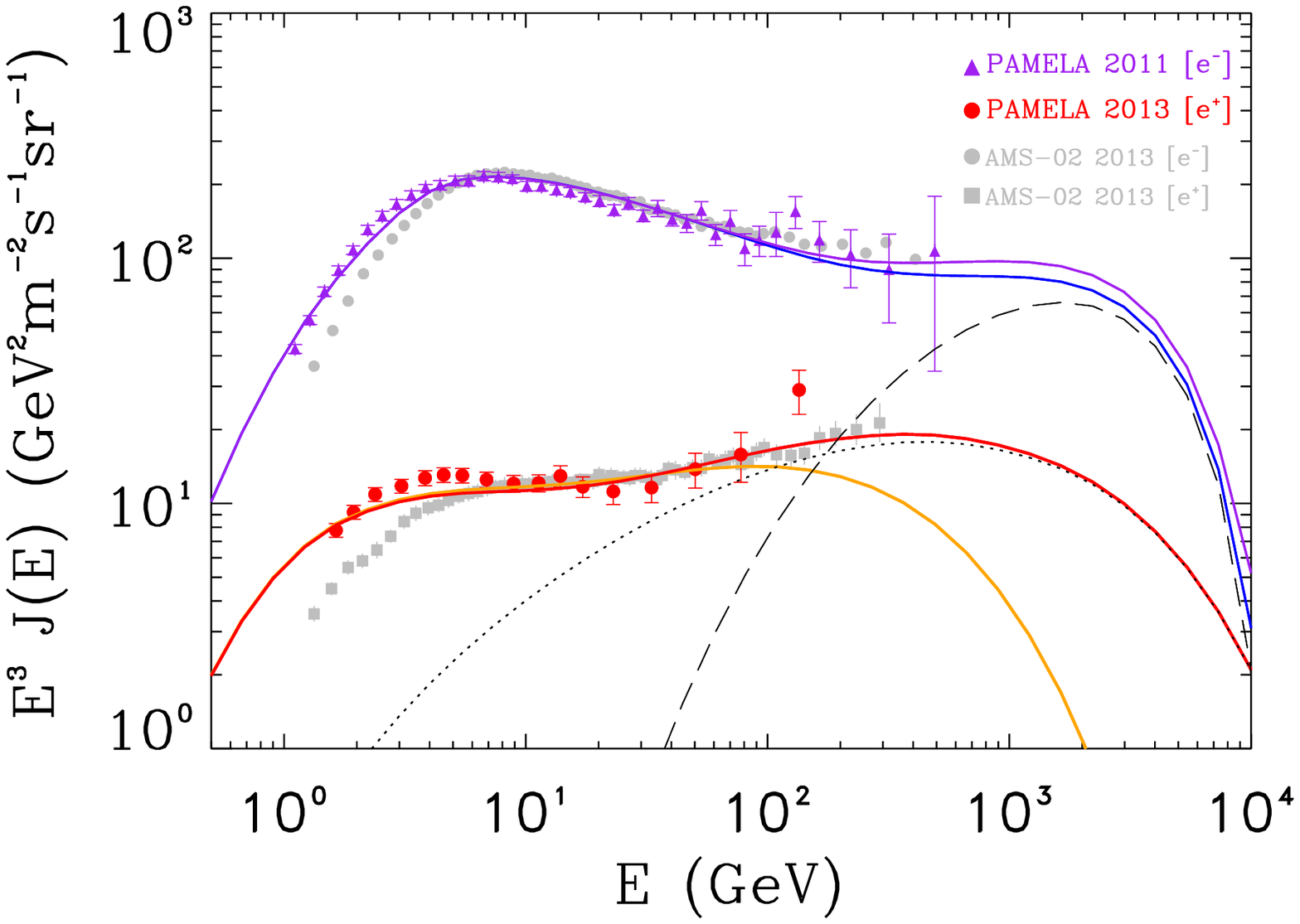}
\includegraphics[width=0.49\textwidth]{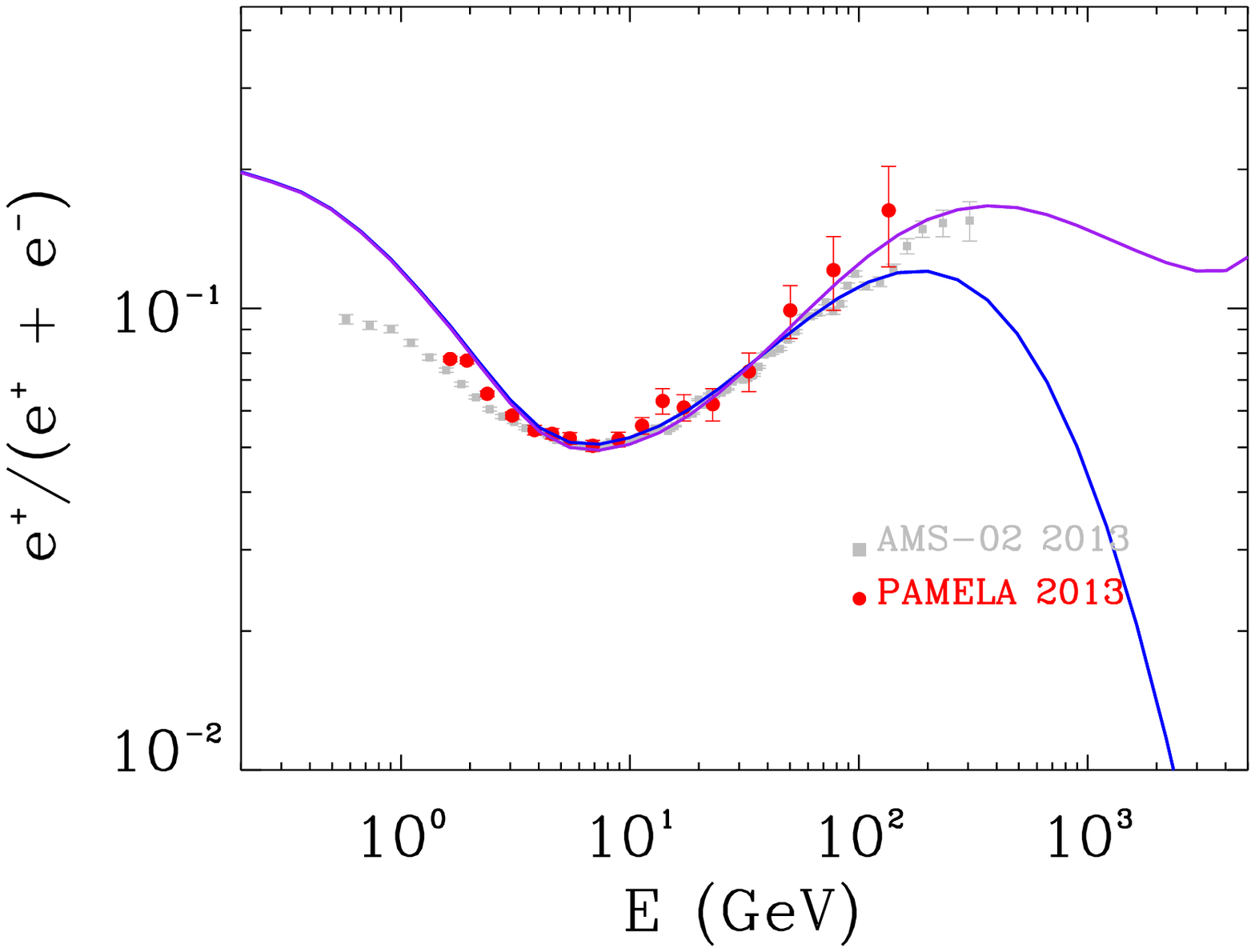}
\caption{The $e^-$ and $e^+$ spectra (upper panel) and the positron fraction (lower panel) obtained by adding to the KRA model components the $e^-$ contribution due to the Vela supernova remnant (dashed line) are shown. The line notation is the same as that of Fig.\ref{fig:E3_spectra}. Only modulated spectra, under the condition suited for the PAMELA data taking period, are shown.}
\label{fig:Vela}
\end{figure*}

From Fig.\ref{fig:Vela} the reader can see how, with respect to the models described in the previous section, the contribution of Vela allows us to improve significantly the agreement between our model and the PAMELA and AMS-02 data above 100 GeV.  The PF data sets are also nicely reproduced in this case.  

\subsection*{B. Adding a $e^- + e^+$ component from nearby pulsars}

Here we consider the possible contribution of Monogem (PSR B0656+14) and Geminga. 

The Monogem distance and age are $d \simeq 280~ \pc$ and $T = 1.1 \times 10^5\ {\rm years}$, respectively, as reported in the ATNF catalog \footnote{http://www.atnf.csiro.au/research/pulsar/psrcat/} \cite{Manchester:2004bp}. 
We estimated its released spin-down energy as was done in Refs.~\cite{Hooper:2008kg,Grasso:2009ma}:
 $E = 1.5 \times 10^{48}\ \erg$.
For Geminga  $d \simeq 160~\pc$,  $T = 3.7 \times 10^5\  {\rm years}$, and  $E = 1.2 \times 10^{49}\ \erg$. 
Similarly to what was done in Refs.~\cite{Grasso:2009ma,DiBernardo:2010is}, we assume an instantaneous $e^\pm$ pairs release into the ISM taking place $5 \times 10^4\  {\rm years}$ after the pulsar birth.
  
\begin{figure*}[t]
\centering
\includegraphics[width=0.49\textwidth]{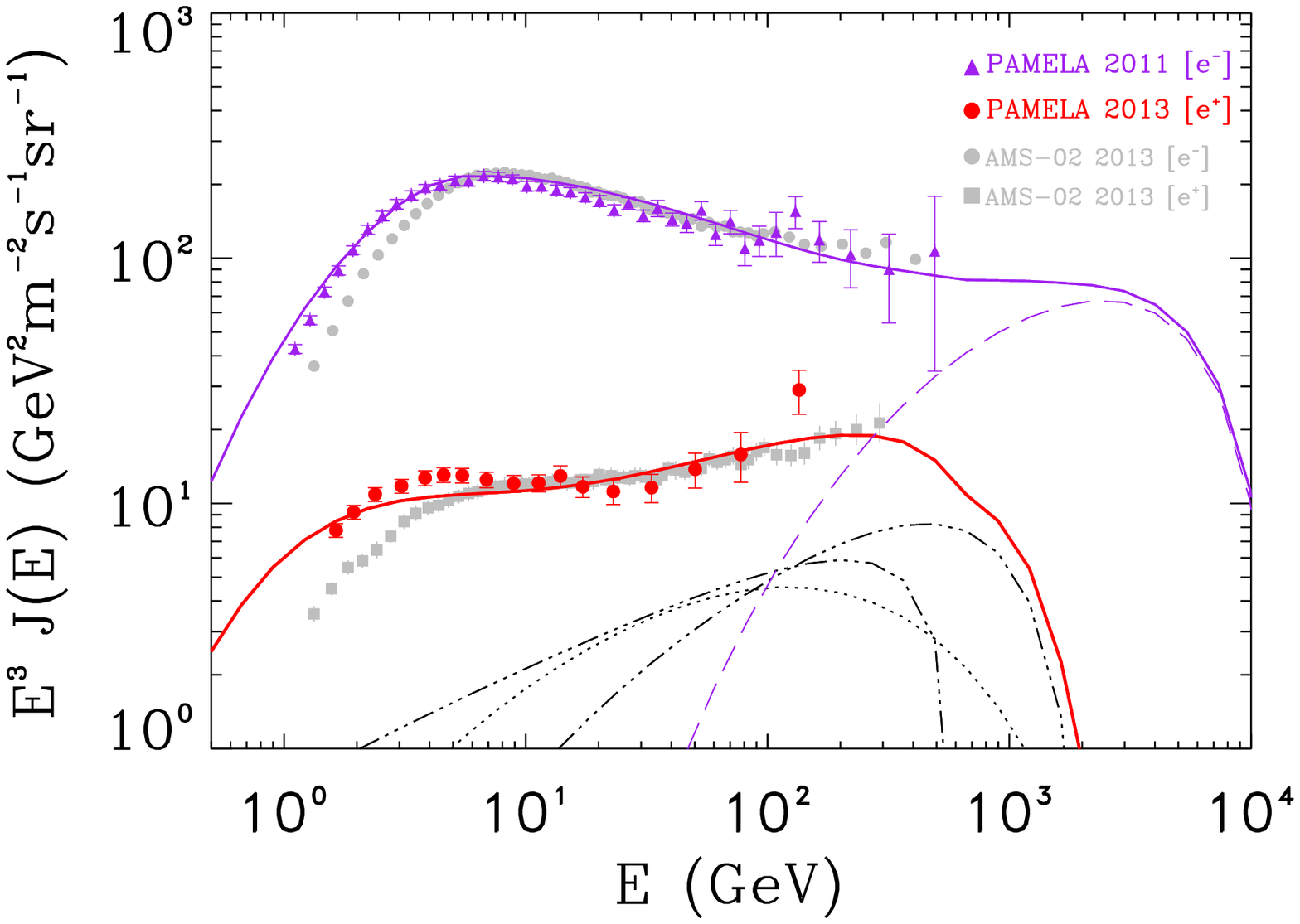}
\includegraphics[width=0.49\textwidth]{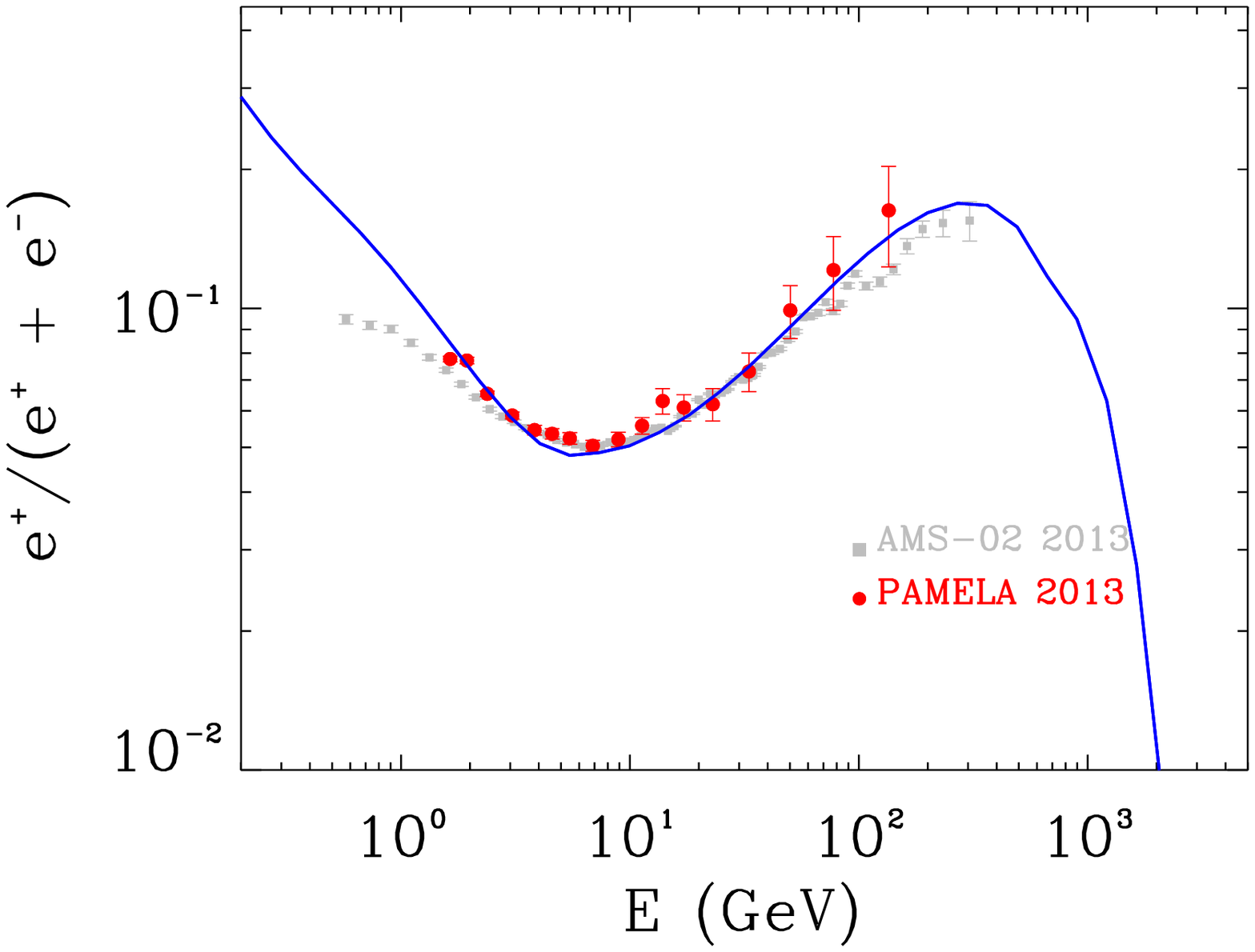}
\caption{The $e^-$ and $e^+$ positron spectra (upper panel) and the positron fraction (lower panel) obtained by adding to the component shown in Fig.~\ref{fig:Vela} the contribution of Monogem and Geminga pulsars (triple-dotted-dashed lines). The cutoffs $E_{\rm cut} = 1~\TeV$ for the Galactic extra component and $E_{\rm PSR} = 1.2~\TeV$ for those pulsars are adopted.
The line notation is the same of Fig.\ref{fig:E3_spectra}.}
\label{fig:pulsars}
\end{figure*}

We find that an injection spectral index $\gamma(e^\pm) = 1.9$ cutoff energy $E_{\rm PSR} = 1.2~\TeV$ for those pulsars provides a very good description of experimental data (see Fig.\ref{fig:pulsars}). 
For consistency we use $E_{\rm cut} = 1~\TeV$ for the extra component sources in the spiral arms (see Fig.\ref{fig:pulsars}).

In this scenario Monogem and Geminga provide a significant fraction of the extra positrons and electrons, on top of a less prominent Galactic contribution.
The important point is that the efficiencies required for these pulsars are $15\%$ and $22\%$, respectively, a factor of 2--20 higher than the values required for Galactic pulsars in the scenario in which they are located in the Galactic arms only. 

\section{Discussion}\label{sec:discussion}

We showed in the above the great relevance of PAMELA results and AMS-02 preliminary data in the leptonic sector both in low- ad high-energy sectors. 

Below 10 GeV, where the $e^+$ spectrum is dominated by the contribution of secondary particles, PAMELA $e^+$ and $e^-$ data provide valuable information on the propagation properties of CR. 
Previous experimental results had a much lower constraining power due to their larger experimental errors and the larger uncertainties related to the treatment of solar modulation.    
The latter is strongly reduced for PAMELA data thanks to the exceptionally low solar activity during which they were taken. In fact, we verified that a detailed treatment of particle propagation in the complex heliosphere magnetic field performed with the {\sc HelioProp} code leads to results very similar to those obtained with the {\it classical} force-field approximation (only in that period though). 
We showed that once the proton injection spectrum is tuned to reproduce PAMELA results the modulated $e^+$ and $e^-$ spectra can consistently be reproduced with DRAGON. 
A good fit of low-energy positrons, however, is possible only for models with low, or null, reacceleration and $\delta \simgeq 0.5$, which adopt a single power law for the proton source spectrum. This result provides a compelling confirmation of previous findings based on PAMELA antiproton data~\cite{DiBernardo:2009ku}. AMS-02 low-energy data are also consistently reproduced for a different {\sc HelioProp} setup. 

At larger energies, our analysis exploits the three-dimensional features of the {\sc DRAGON} code to properly account the effect of energy losses in the presence of a realistic spiral-arm distribution of astrophysical sources. 
As shown for the first time in Ref.~\cite{Gaggero:2013rya}, such an approach reveals that, above 10 GeV, the $e^+$ and $e^-$ spatial distribution in the Galaxy should be rather inhomogeneous, and their local spectrum should be significantly steeper than that obtained with azimuthally symmetric (two-dimensional) numerical or semianalytical computations. Here we confirm those findings. We notice, however, that PAMELA data still point to a rather steep source spectrum of the $e^-$ background (see Table~\ref{tab:prop_para}) with respect to that expected from Fermi acceleration. This may suggest that rather strong energy losses occur before the escape of the CR electrons from the acceleration region.  

The other relevant issue studied in this paper concerns the features, and the nature, of the electron and positron extra component required to explain the positron fraction rise above 10 GeV. 
Here we showed that the $e^+$ and $e^-$ absolute spectra measured by PAMELA and AMS-02 also requires this new component. 
Both data sets can be consistently, and accurately, reproduced if the extra component sources are distributed along the Galaxy spirals arms, as expected both for SNR and pulsars and as we assumed for the CR primary nuclei and the $e^-$ background sources. 

For the first time, in this work we determined the injection power required for the Galactic extra component. 
We argue that both a mechanism of secondary production at the acceleration site as proposed in Ref.~\cite{Blasi:2009hv} and the pair production of leptons in the magnetospheres of the pulsars~\cite{Hooper:2008kg} may satisfy the energy requirements. In fact, the energy injection rate needed to sustain the observed extra flux is about 4 orders of magnitude lower than the total injected power from supernovae and about 2 orders of magnitude lower than the total injected power from pulsars.
Therefore, other observables should be considered in order to discriminate between those two scenarios. 

In agreement with previous results (see Ref.~\cite{Serpico:2011wg} and references therein), the source spectrum of such a Galactic extra component must be rather hard $(\gamma(e^-) \simeq 1.7)$. We showed that PAMELA data do not allow us to discriminate between relatively low values of the extra component cutoff $E_{\rm cut} \simeq 1~ \TeV$ from larger values. This is a very important issue since a value of $E_{\rm cut} \gg 1~ \TeV$ would disfavor a pulsar origin of this component and be also rather challenging for dark matter interpretations. 

Although PAMELA data are compatible with a charge symmetric extra component, from Fig.\ref{fig:E3_spectra}  we notice that in that case our models slightly underpredict the $e^-$ flux above few hundred GeV.  
This tension is much more severe if AMS-02 preliminary data are considered.   
Such a discrepancy may point to a charge asymmetry in the extra component production mechanism or to the presence of nearby $e^-$ accelerators.  
Furthermore, we notice that both $e^+$ and $e^-$ AMS-02 preliminary data clearly disfavor a purely Galactic origin of the extra component if $E_{\rm cut} \simeq 1~ \TeV$. 
This is opposite of what was inferred on the basis of the AMS-02 positron fraction alone \cite{Linden:2013mqa,Cholis:2013psa,Gaggero:2013rya}.  

In Sec.\ref{sec:local_sources} we considered the effect of nearby sources on our results. We assumed that propagation in the local region behaves as in the rest of the Galaxy (which we are aware may not be true \cite{Kistler:2012ag}). We found the following:\\
{\rm a)} If the $e^\pm$ extra component originates in the Galactic arms with a high-energy cutoff ($E_{\rm cut} \simeq 10~ \TeV$),
the presence of a $e^-$ contribution coming from a nearby SNR ({\rm e.g.} Vela) improves the match between our model and PAMELA and AMS-02 data. \\ 
{\rm b)} For a lower-energy cutoff ($E_{\rm cut} \simeq 1~ \TeV$), the presence of a significant $e^+ + e^-$ local component (on top of the Galactic extra component) allows us to reproduce AMS-02 preliminary data. Observed nearby pulsars are viable source candidates for such local components although the required pair conversion efficiency is rather high (order of 20\%). 

Even more extreme cases in which the bulk of extra electrons and positrons originates from few nearby pulsars only, with even higher efficiency, appear less convincing because almost half of the total spin-down energy should go in the pair production process. Moreover, there is no reason why such energetic objects should be located in the solar vicinity only.
Therefore, the intermediate scenario, where the extra component is originated by a population of sources along the arms plus some contribution from few energetic nearby pulsars, turns out to be more appealing (see also Ref.~\cite{DiMauro:2014iia} where, however, the arm distribution of distant sources was not considered).  

We remind the reader that an interpretation of the positron CR excess in terms of particle dark matter annihilation or decay has been proposed many times in the literature (see, e.g., Ref.~\cite{Cirelli:2012tf} for a recent review). 
Although these models are quite challenging from the model building point of view (see, e.g., Ref.~\cite{ArkaniHamed:2008qn}) an interpretation of available data in those term is still viable. 
We notice, however, that a consistent treatment of the astrophysical background and the DM annihilation/decay product propagation is rarely performed. 
Furthermore, a realistic electron and positron background computed accounting for the spiral-arm distribution of astrophysical source was not adopted so far. 
This may have a significant impact on the DM indirect search constraints. 
On the basis of our previous considerations, we also argue that recent DM models which have been tuned against PAMELA and AMS-02 positron fraction results may be severely constrained by the $e^+$ absolute spectrum measured by PAMELA and AMS-02.         
 
{\em Acknowledgments:}  We thank I.~Gebauer  for valuable discussions.
This research was partly conducted using the resources of High Performance Computing Center North (HPC2N). 
C.~E. acknowledges support from the ‘‘Helmholtz Alliance for Astroparticle Physics HAP’’ funded by the Initiative and Networking Fund of the Helmholtz Association.

\bibliographystyle{apsrev4-1}
\bibliography{bibliography}

\end{document}